\def\ra{\rangle}
\def\la{\langle}
\def\be{\begin{equation}}
\def\ee{\end{equation}}
\def \qed { \mbox{}\hfill $\Box$\vspace{1ex}}

\def\ee{\end{equation}}
\def\ba{\begin{array}}
\def\ea{\end{array}}

\def\Cb{{\Bbb C}}

\documentstyle[12pt]{article}
\input amssym.def
\begin{document}
\baselineskip18pt \thispagestyle{empty}
\begin{center}
{\LARGE \bf  Separability and Entanglement of Identical Bosonic Systems}
\end{center}
\vskip 2mm

\begin{center}
{\normalsize Xiao-Hong Wang$^1$,  Shao-Ming Fei$^{1,\  2}$  and Ke Wu$^1$ }
\end{center}

\begin{center}
\begin{minipage}{5.3in}
{\small \sl $^1$ Department of Mathematics, Capital  Normal
University, Beijing, China}

{\small \sl $^2$ Max-Planck-Institute for Mathematics in the Sciences, 04103 Leipzig, Germany}

\end{minipage}
\end{center}

\begin{center}
\begin{minipage}{5.2in}
\vskip 3mm {\bf Abstract}
We investigate the separability
of arbitrary $n$-dimensional multipartite identical bosonic systems.
An explicit relation between the dimension and the separability
is presented. In particular, for $n=3$, it is shown that
the property of PPT (positive partial transpose)
and the separability are equivalent for tripartite systems.

\vskip 1mm
PACS number(s): 03.67.Hk, 03.65.Ta, 89.70.+c
\end{minipage}
\end{center}

\bigskip
\bigskip

Quantum entanglement plays essential roles in quantum information
processing and quantum computation. The entangled states provide
key resources for a vast variety of novel phenomena such as
quantum cryptography, quantum teleportation, super dense coding,
etc \cite{nielsen}. An important problem in the theory of quantum
entanglement is the separability. One of the famous separability
criterion was given by Peres \cite{peres96}. It says that all
separable states necessarily have a positive partial transpose
(PPT), which is further shown to be also sufficient for states on
$\Cb^2 \otimes \Cb^2$ and $\Cb^2 \otimes \Cb^3$
\cite{horo96,tran}, where $\Cb^n$ denotes the $n$-dimensional
complex space. There have been many results on the separability
and entanglements of mixed states, see e.g.,
\cite{1,22n,23n,we,prl}. In particular, it is shown that every
quantum states $\rho$ supported on $\Cb^M \otimes \Cb^N$, $\Cb^2
\otimes \Cb^2 \otimes \Cb^N$ and $\Cb^2 \otimes \Cb^3 \otimes
\Cb^N$ with positive partial transposes and rank $r(\rho)\leq N$
are separable and have a canonical form \cite{1,22n,23n}.

Although the entanglement is extensively studied for distinguishable
particle systems, the entanglement of identical particle systems has
been less investigated.
In fact in certain systems such as quantum dots \cite{SLM:00},
Bose-Einstein condensates \cite{r2} and parametric down
conversion \cite{r3}, the entanglement should be treated as the one
of identical particle systems. Schliemann et
al \cite{SLM:00,SCK+:00} have discussed the entanglement in
two-fermion systems. They found that the entanglement in two-fermion
systems is analogous to that in a two-distinguishable particle
system. The results
for two-boson systems are quite different. Li {\it et al.}
\cite{LZLL:01} and Paskauskas and You \cite{PaYo:01} have studied this
problem of two-boson systems.
For multipartite bosonic systems, there are very few discussions. Recently,
the author in \cite{ZZhXY05} obtained the canonical form for pure
states of three identical bosons, and classified the
entanglement correlation into two types, the analogous GHZ and the
W-types. In \cite{ESBL02}, it has been shown that rank
$n$ and rank $\frac{n(n+1)}{2}-2$ PPT bosonic mixed states in the
symmetrized tensor product space ${\cal S}(\Cb^n\otimes \Cb^n)$ are separable, and all
three-qubit ($n=2$) bosonic PPT states are separable as well.
For bosonic mixed
state $\rho$ in $k-$qubit system, $k\geq 4$, $\rho$ is $PPT$ implies
that $\rho$ is separable, except for the case of maximal rank.

In this letter, we investigate the separability
of multi-partite identical bosonic systems with arbitrary dimension $n$.
Let  ${\cal H}={\cal S}(\Cb^{n}\otimes \Cb^{n} \otimes \cdots \otimes\Cb^{n})$ denote the symmetrized tensor product space of $k$
$n$-dimensional spaces associated with Alice, Bob, Charle, etc.
The dimension of the space ${\cal H}$ is given by \cite{formula},
\be\label{Bosdim}
I^k_n=\frac{(n+k-1)!}{k!(n-1)!}=C^k_{n+k-1}.
\ee

We first consider the case of $k=3$.

\noindent{\bf [Theorem 1]} {\rm Let $\rho$ be a bosonic mixed state
 in ${\cal S}(\Cb^{n} \otimes\Cb^{n} \otimes\Cb^{n})$,
with a positive partial transpose with respect to Alice.
If the rank of $\rho$, $r(\rho)\leq n^2$, then $\rho$ is separable.}

\noindent{\bf Proof.} We first prove the case of $n=3$.  Suppose
that the state $\rho$ is a PPT state with respect to Alice and has
a rank 9. We can treat it as a bipartite PPT state in a $3\times
9$ dimensional space of Alice-(Bob,Charlie). From the Theorem 1 in
\cite{1} ( also Theorem 1 in \cite{22n}), such a state of rank 9
is necessarily separable and can be represented as
$\rho=\sum_{i=1}^9p_i| e_i,\Psi_i\rangle\langle e_i,\Psi_i|$,
where the vectors $|\Psi_i\rangle$ are generally entangled pure
states associated with the spaces of Bob and Charlie. As
$|\Psi_i\rangle$ are mutually orthogonal, they belong to the range
of the reduced density matrix (partial trace with respect to the
space associated with Alice) ${\rm Tr}_A\rho$, and hence
$|\Psi_i\rangle \in {\cal S}(\Cb^{3} \otimes\Cb^{3})$. Moreover
$|e_i,\Psi_i\rangle$ belong to the range of $\rho$. Therefore
$|e_i,\Psi_i\rangle\,\in {\cal S}(\Cb^{3}
\otimes\Cb^{3}\otimes\Cb^{3})$. According to Schmidt decomposition
we can write $|\Psi_i\rangle=
a_i|00\rangle+b_i|11\rangle+c_i|22\rangle$ for some $a_i, b_i,
c_i\in \Cb$, where $|0\ra, |1\ra, |2\ra,$ are the Schmidt basic
vectors in $\Cb^{3}$. The only possible forms of
$|e_i,\Psi_i\rangle$ satisfying the above conditions are
$|000\rangle$, $|111\rangle$ or $|222\rangle$. Therefore $\rho$ is
separable.

When the rank of $\rho$ is strictly less then 9, $\rho$ can be embedded
into a smaller space. For instance, if $r(\rho)=8$, $\rho$ is
supported on spaces $2\times 8$ or $3\times 8$. $\rho$ is then separable
in the partition Alice-(Bob,Charlie) and can be again written as
$\rho=\sum_{i=1}^8p_i| e_i,\Psi_i\rangle\langle e_i,\Psi_i|$. By
using the same procedure as above, we can prove that
$| e_i,\Psi_i\rangle$ is fully separable, and hence $\rho$ is separable.
The general $n$-dimensional case can be proved similarly. \qed

\noindent{\bf [Remark ]} From the theorem we see that a bosonic mixed state $\rho$
in ${\cal S}(\Cb^{3} \otimes\Cb^{3} \otimes\Cb^{3})$
with a positive partial transpose is separable if
$r(\rho)\leq 9$. As the
dimension of the space of  ${\cal S}(\Cb^{3} \otimes\Cb^{3} \otimes\Cb^{3})$ is 10, Theorem 1 says that
almost all the PPT bosonic mixed states in
${\cal S}(\Cb^{3} \otimes\Cb^{3}
\otimes\Cb^{3})$ are separable, except for the case
$r(\rho)=10$. Hence the
rank of a bound entangled state in ${\cal S}(\Cb^{3} \otimes\Cb^{3} \otimes\Cb^{3})$ has to be 10.

When $n=4$, we have $I^3_4=20$. As $\rho$ is separable if $r(\rho)\leq 16$,
all bound entangled states $\rho$ in ${\cal S}(\Cb^{n}\otimes\Cb^{n}
\otimes\Cb^{n})$ satisfy  $17\leq r(\rho) \leq 20$.

\noindent{\bf [Theorem 2]} {\rm Let $\rho$ be a PPT bosonic mixed
state in ${\cal S}(\Cb^{n}
\otimes\Cb^{n} \otimes \cdots \otimes\Cb^{n})$ with $k$
subsystems ($k\geq 4$). If $r(\rho)\leq I^{k-1}_n$, then $\rho$ is
separable.}

\noindent{\bf Proof.} We prove the case of $n=3$ ( the
other cases can be proved similarly). Assume that $\rho$ is
PPT, say with respect to the space associated with Alice,
with rank $I^{k-1}_3=\frac{k(k+1)}{2}$.

If we consider $\rho$ as a bipartite state in the partition
Alice - the rest, $\rho$ is supported on $\Cb^{3}\otimes{\cal S}((\Cb^{3})^{\otimes k-1})$. From
\cite{1} $\rho$ is separable with respect to this
partition and has a form,
$\rho=\sum_{i=1}^\frac{k(k+1)}{2} p_i| e_i,\Psi_i\rangle\langle
e_i,\Psi_i|$, where $|e_i\rangle$ (resp. $|\Psi_i\rangle$)
are vectors on the spaces associated to Alice (resp. the rest).

We prove result by induction. We illustrate the
procedure by proving the case of $k=4$. As $|\Psi_i\rangle$ belong to the
range of the reduced density matrix ${\rm Tr}_A\rho$, they must
belong to ${\cal S}(\Cb^{3}\otimes\Cb^{3}\otimes\Cb^{3})$. Since $\rho$ is PPT $|\Psi_i\ra\la\Psi_i|$ is a PPT state in ${\cal
S}(\Cb^{3}\otimes\Cb^{3}\otimes\Cb^{3})$.
However the rank $r(|\Psi_i\ra\la\Psi_i|)=1$,
from Theorem 1, $|\Psi_i\ra$ is separable, and can written as
$|\Psi\ra=|f_i, f_i,f_i\ra$ for some vectors $|f_i\ra$ in $\Cb^{3}$. While the vectors $|e_i,\Psi_i\rangle$ belong to the
range of $\rho$ and hence
$|e_i,\Psi_i\rangle \in {\cal S}(\Cb^{3}\otimes\Cb^{3}\otimes\Cb^{3}\otimes\Cb^{3})$. Therefore the only possible forms of
$|e_i,\Psi_i\rangle$ are $|f_i, f_i, f_i, f_i\rangle$. Therefore
$\rho$ is separable. \qed

We have presented some separability criteria for
multipartite bosonic mixed states. For tripartite PPT states,
all bound entangled states have necessarily rank greater than $n^2$.
For general multipartite PPT bosonic states with $k$
subsystems ($k\geq 4$), if $r(\rho)\leq I^{k-1}_n$, $\rho$ is
separable. The results can be used to construct possible bound
entangled states of identical bonsonic systems.
For instance, if $k=4$, $n=3$, we have $I^4_3=15$. The rank
of a bound entangled state has to be between $I^3_3=10$ and $15$.

\end{document}